\pdfoutput=1

\documentclass[11pt,twoside,a4paper,cmspaper,final,collab]{cms-tdr}
\def\svnVersion{58567:58620}\def\cmsCernNo{2011-053}\def\cmsCernDate{\today}\def\cmsMessage{Submitted to Physical Review Letters}

\begin{document}\cmsNoteHeader{QCD-10-011}

\hyphenation{had-ron-i-za-tion}
\hyphenation{cal-or-i-me-ter}
\hyphenation{de-vices}
\RCS$Revision: 58620 $
\RCS$HeadURL: svn+ssh://alverson@svn.cern.ch/reps/tdr2/papers/QCD-10-011/trunk/QCD-10-011.tex $
\RCS$Id: QCD-10-011.tex 58620 2011-06-01 15:26:59Z alverson $
\providecommand{\HERWIGPP} {{\textsc{herwig++}}\xspace}
\providecommand{\NLOJETPP} {{\textsc{nlojet++}}\xspace}

\cmsNoteHeader{QCD-10-011} % This is over-written in the CMS environment: useful as preprint no. for export versions
\title{Measurement of the Inclusive Jet Cross Section
 \texorpdfstring{\\ in \Pp\Pp\ Collisions at $\sqrt{s} = 7$~TeV}{ in pp Collisions at sqrt(s) = 7 TeV}}% Force line breaks with \\

\ifthenelse{\boolean{cms@external}}
{%
}
{%
}

\date{\today}

\abstract{
The inclusive jet cross section is measured in \Pp\Pp\ collisions with a center-of-mass energy of 7 TeV at the LHC using the CMS experiment. The data sample corresponds to an integrated luminosity of 34~${\rm pb^{-1}}$. The measurement is made for jet
transverse momenta
in the range 18--1100~GeV and for
absolute values of rapidity less than 3.
The measured cross section extends to the highest values of jet $p_{\rm T}$ ever observed and, within the experimental and theoretical uncertainties, is generally in agreement with next-to-leading-order perturbative QCD predictions.
}

\hypersetup{%
  pdfauthor={CMS Collaboration},%
  pdftitle={Measurement of the Inclusive Jet Cross Section in pp Collisions at sqrt(s) = 7 TeV},%
  pdfsubject={CMS},%
  pdfkeywords={CMS, physics, QCD}
}
\maketitle %maketitle comes after all the front information has been supplied

\def\verALL{1}
\def\verPF{2}
\def\ver{1}

\clearpage

The measurement of the inclusive jet cross section is a benchmark of the standard model (SM) at hadron colliders~\cite{Rohlf1,Rohlf2}. At the Large Hadron Collider (LHC), jets produced in the high center-of-mass energy collisions test the SM at the smallest distance scales presently possible, can constrain parton momentum distributions in the proton, and are sensitive to the strong coupling constant.  Significant deviations from predictions of the inclusive jet cross section at high transverse momentum \pt could also be an indication of new phenomena beyond the SM.  Results from the Tevatron \Pp\Pap\ collider demonstrate agreement with next-to-leading-order (NLO) theoretical predictions from perturbative Quantum Chromodynamics
(pQCD) for jets in the \pt range of 50--650~GeV, using approximately
1 fb$^{-1}$ of data at a center-of-mass energy $\sqrt{s} = 1.96$~TeV~\cite{Aaltonen:2008eq, Abazov:2008hu, Abulencia:2007ez}.
Early results from the ATLAS Collaboration for jets in the \pt range 60--600~\GeV, based on a 17\nbinv data sample of \Pp\Pp\ collisions at $\sqrt{s}=7$~\TeV at the LHC, also indicate agreement with theoretical predictions~\cite{Collaboration:2010wv}.
Using a data sample corresponding to $34$~pb$^{-1}$ of integrated luminosity from \Pp\Pp\  collisions recorded by the CMS detector at the LHC with $\sqrt{s}=7$~\TeV, we significantly extend the \pt range from previous measurements of the inclusive jet \pt spectrum to 18--1100~\GeV and for rapidities $|y| < 3.0$.
The rapidity $y$ is defined as $y = \frac{1}{2}\ln[(E+p_z)/(E-p_z)]$, where $E$ is the jet energy and $p_z$ is the component of the jet momentum along the beam axis.
In this Letter, the inclusive jet cross section, differential in \pt and $y$, is compared with theoretical predictions at NLO in pQCD.

The CMS detector has silicon pixel and microstrip trackers covering pseudorapidities up to $|\eta| = 2.5$,
where $\eta = -\ln[\tan(\theta/2)]$ and $\theta$ is the polar angle relative to the counterclockwise proton beam direction. Together with a 3.8 T solenoid, the trackers provide track reconstruction down to transverse momenta of about 100\,\MeV and a resolution of about 1\% at 100\,\GeV. A high-granularity electromagnetic crystal calorimeter (ECAL)
extends up to $|\eta| = 3.0$, and has an energy resolution of better than 0.5\,\% for unconverted photons with transverse energies above 100\,GeV.
A hermetic hadronic calorimeter (HCAL) extends up to $|\eta| = 5.0$
with a transverse hadronic energy resolution of about $100\,\%/\sqrt{E_{\rm T}\,[\GeV]} \oplus 5\,\%$. The calorimeter components relevant to this work may be described in terms of a cylindrical barrel region, extending up to $|\eta| = 1.5$, and two endcaps, covering $1.5 < |\eta| < 3.0$. An efficient muon system reconstructs and identifies muons up to $|\eta| = 2.4$.
Events are collected using a two-level trigger system, consisting of a hardware Level-1 and a software high level trigger. Jets formed online by the trigger system use the energies measured in the ECAL and HCAL, and are uncorrected for the jet energy response of the calorimeters. This study uses inclusive single-jet triggers corresponding to \pt thresholds of 6, 15, 30, 50, 70, 100, and 140~GeV. Finally, a minimum-bias trigger is defined as a signal from at least one of two beam scintillator counters in coincidence with a signal from one of two beam pick-up timing devices.
The CMS coordinate system is right-handed, with the origin centered at the nominal collision point, the $x$~axis pointing radially toward the center of the LHC, the $y$~axis pointing vertically upward, and the $z$~axis pointing along the beam direction.
More details about the CMS detector can be found in Ref.~\cite{CMSExperiment}.

This measurement uses the infrared- and collinear-safe anti-$k_T$ jet clustering algorithm~\cite{ANTIKT}, as implemented in Ref.~\cite{Cacciari:2005hq}. A distance parameter of $R=0.5$ is chosen.
The CMS particle-flow (PF) algorithm reconstructs individual particles (leptons, photons, charged and neutral hadrons) by linking tracks, ECAL clusters, and HCAL clusters. The momentum or energy of each particle is measured based on information from all subdetectors. Broadly speaking, electrons are reconstructed from tracks and calibrated ECAL clusters; muons are reconstructed using tracks; charged hadrons are reconstructed from tracks and calibrated ECAL and HCAL clusters; photons and neutral hadrons are reconstructed from calibrated ECAL and HCAL clusters, respectively.  A detailed description of the PF algorithm and performance can be found in Ref.~\cite{PFT-10-002} and references therein. Jets are reconstructed by clustering particles, including leptons.
Because the energy response of each reconstructed particle is close to unity, the jet momentum response is also close to unity and only small corrections (at the level of 5--10\%) are needed~\cite{CMS-PAS-JME-10-014, JME-10-010}. The bulk of these corrections are derived from simulation \cite{GEANT4} with residual corrections determined from data. First, using an exclusive dijet sample containing back-to-back jets in azimuth, a residual relative correction of up to 8\% removes any remaining pseudorapidity dependence of the jet momentum response. Second, using an exclusive data sample containing a single hadronic jet recoiling back-to-back in azimuth against a well measured photon \cite{JME-10-010}, a small residual absolute correction of about 1\% restores the response to unity.
Jet momenta are further corrected for the pile-up of multiple proton-proton collisions~\cite{Cacciari:2007fd}, which ranged from nearly zero additional collisions at the very early period of LHC data taking in 2010 to an average of about three near the end of the 2010 running period.
Finally, the jet momentum resolution is determined as a function of the jet \pt and $y$ from simulation. Comparing the \pt balance in dijet events between data and simulation, the jet \pt resolution from simulation is scaled to be about 10\% larger in the barrel and about 20\% larger in the endcaps to match that from data~\cite{CMS-PAS-JME-10-014, JME-10-010}.

The efficiency of a given inclusive single-jet trigger is determined as the ratio of its trigger rate to either the minimum-bias trigger rate or the rate of the lower-$\pt$ trigger preceding it. The minimum jet \pt is chosen so that the trigger efficiency exceeds 99\% for each dataset.
All events are required to have a primary vertex (PV) satisfying the following selection: the fit for the PV must include at least three associated tracks, the PV must lie within $0.15~\cm$ of the beam axis, and the $z$~coordinate of the PV must lie within the luminous collision region $|z_{\rm PV}| < 24~\cm$.
This selection rejects noncollision and beam-related backgrounds.
Loose jet identification criteria~\cite{CMS-PAS-JME-10-001} are applied requiring that each jet within the tracker acceptance have at least
two particles, of which at least one must be a charged hadron.
Further, at most 90\% of the jet energy is allowed to be from photons or neutral hadrons. Beyond the tracker acceptance, each jet is required to have both electromagnetic and hadronic energy. To remove effects from anomalous calorimeter noise, beam halo, or cosmic-ray backgrounds, events are rejected if the missing transverse momentum is both larger than 50\% of the total visible transverse energy and larger than 100~GeV. The missing transverse momentum is defined as the modulus of the negative transverse vector sum over the momenta of all reconstructed particles in the event. Any inefficiency in selecting jets, due to the above criteria, is estimated from simulation to be negligible.

Due to resolution effects,
a jet may fall into a different \pt bin than the one corresponding to the
true underlying jet (unaltered by experimental effects).
Such bin-to-bin migrations distort the rapidly falling \pt spectrum.
Each \pt bin width is chosen to be larger than the reconstructed jet \pt resolution in the bin to minimize the migration effects and large enough to ensure that statistical fluctuations do not dominate the measurement.  The jet \pt~spectra are corrected for resolution effects where the true jet $\pt$~spectrum is modeled by a %four-parameter modified power-law function of the jet transverse momentum
power-law {\it ansatz} motivated by the parton model,
modified by a kinematic cutoff term at high \pt:
$f(\pt; \alpha,\beta,\gamma)~=~N_0\big[\pt\big]^{-\alpha}\big[1~-~\frac{1}{\sqrt{s}}2\pt\cosh(y_{\rm min})\big]^\beta$ $\exp[-\gamma/\pt]$, %\nonumber
where $N_0$ is a normalization factor, $\alpha$, $\beta$, $\gamma$ are fit parameters, and $y_{\rm min}$ is the low-edge of the rapidity bin $y$ under consideration.
Similar parameterizations have been previously used by other experiments~\cite{CDF-ANSATZ,Abazov:2008hu,Collaboration:2010wv}.
The function is then smeared using the jet \pt resolutions in bins of \pt and $y$.
The parameters of the model are extracted by fitting the smeared transverse momentum spectrum to the experimental data.  The data points are placed at the bin center, defined as the point where the value of the predicted function is equal to its mean value over the bin width.
The \pt resolution corrections for each bin in \pt and $y$, determined by taking the ratio of the function to the smeared function, range from 5 to 10\% as a function of \pt in the central rapidity bin and range from about 10\% to 50\% in the most-forward rapidity bin.
Jet migrations across $y$ bins due to $y$ resolution are found to be negligible within the tracker acceptance.  Using simulation, migrations of up to 5\% are observed and corrected across bins whose boundaries lie near the tracker acceptance.

The primary sources of systematic uncertainties in the cross section measurement arise from the jet momentum scale and resolution, as well as the integrated luminosity.  The jet transverse momentum scale is sensitive to several effects including (a) the photon energy scale (known to 1.0\%~\cite{EGM-10-003} and used to derive the residual absolute response corrections), (b) the relative response across detector regions (known to within 2.6\%~\cite{JME-10-010}), (c) pile-up effects
(known to within 0.2\% for very low pile-up conditions with low-\pt jet triggers or for intermediate pile-up conditions with high-\pt jet triggers),
and (d) the calibration extrapolation to transverse momenta above available photon energies (dominated by uncertainties in jet fragmentation and estimated to be within 4\% at 1100 GeV~\cite{JME-10-010}).  With those considerations, the total uncertainty in the jet transverse momentum scale is determined to be between 3\% and 4\% in the ranges $18 < \pt < 1100\GeV$ and $|\eta| < 3.0$.
The jet momentum resolutions for different $y$ bins are
known to within 10\% at $|y|<1.5$, increasing to 15\% for $1.5<|y|<2.0$, 25\%~for~$2.0<|y|<2.5$, and 30\% for $2.5<|y|<3.0$~\cite{CMS-PAS-JME-10-014}.
The integrated luminosity of the proton--proton collisions is %monitored by two forward hadron calorimeters (HF), with a coverage of $3 < |\eta| < 5$. Vertexing multiple tracks within the interaction region provides an independent luminosity measurement cross-check and is found to be compatible with that from the HF.  Knowledge of the LHC beam currents and a precise study of the transverse size of the beams provides
known with a precision of 4\%~\cite{CMS-DP-2011-002}
and directly translates into a 4\% normalization uncertainty on the inclusive jet cross section.

\begin{figure}[htbp]
  \centering
    \centering
    \includegraphics[width=0.99\columnwidth]{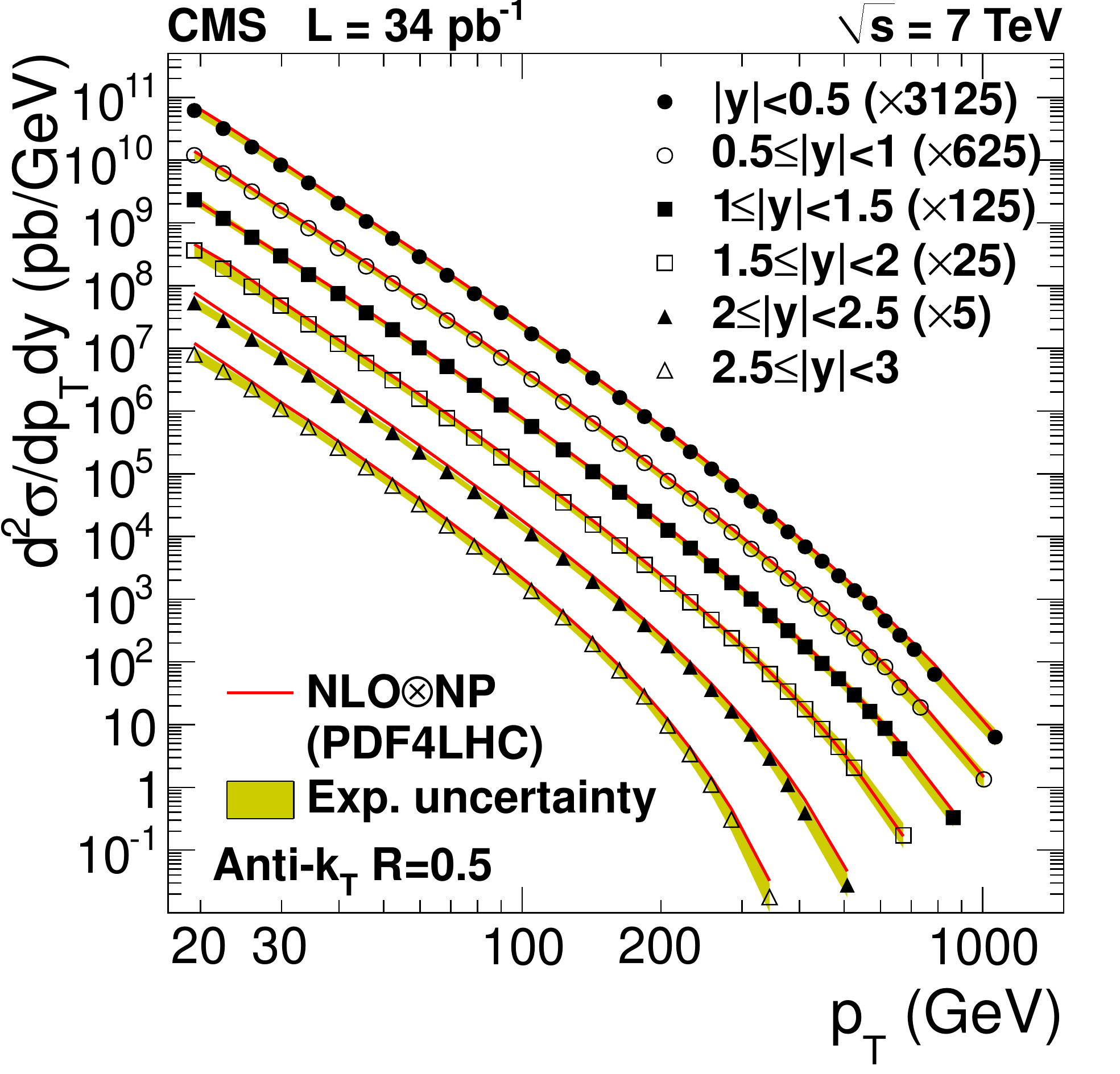}%
    \caption{Fully-corrected inclusive jet differential cross sections as a function of \pt for six different rapidity intervals, scaled by the factors shown in the legend for easier viewing.  The next-to-leading-order (NLO) theoretical predictions, corrected for non-perturbative (NP) effects via multiplicative factors, are superimposed.    }    %
    \label{fig:UnsmearedSpectra}
\end{figure}
\begin{figure*}[htbp]
  \centering
    \centering
    \includegraphics[width=0.95\textwidth]{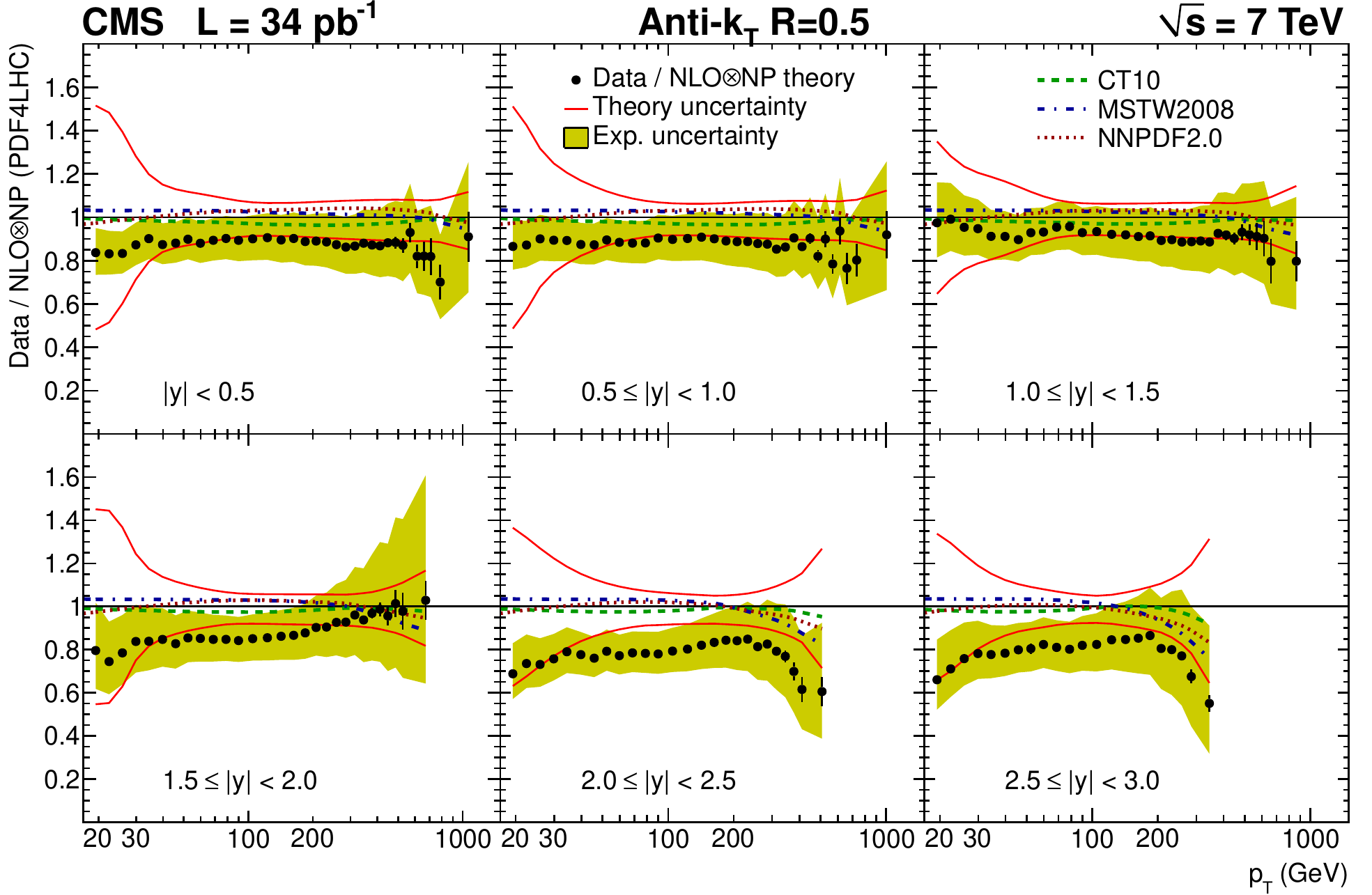}%
    \caption{Ratios of the fully-corrected measured jet \pt differential cross sections to the theoretical prediction as a function of \pt.  The error bars show the experimental statistical uncertainties. The shaded band about the data points represent the total experimental systematic uncertainty.  The solid lines represent the total theoretical systematic uncertainty.  The central predictions for the CT10 (dashed line), MSTW2008NLO (dash-dotted line), and NNPDF2.0 (dotted line) PDF sets are also shown.
    }    %
    \label{fig:UnsmearedRatio}
\end{figure*}

The next-to-leading-order perturbative QCD theoretical predictions are derived using \NLOJETPP~2.0.1~\cite{Nagy:2001fj,Nagy:2003tz} within the framework of {\sc fastNLO}~1.4~\cite{Kluge:2006xs}. The {\sc fastNLO} framework is used for propagating uncertainties due to different parton distribution function (PDF) sets, $\alpha_s$ values, and scale choices.
Non-perturbative (NP) corrections for hadronization and multiple parton interactions are estimated using \PYTHIA~6.422~\cite{Sjostrand:2006za} and \HERWIGPP~2.4.2~\cite{Bahr:2008pv}, which are applied to the NLO pQCD prediction.
The correction is defined as the average of the models, and the associated
theoretical uncertainty is assumed to be half of the difference between the two predictions. For low-\pt jets, the NP correction can be as large as 30\%, with a relative uncertainty of 100\%.
Uncertainties from any residual dependence on the choice of renormalization scale $\mu_{r}$ and factorization scale $\mu_{f}$ are determined by varying the scales according to the following combinations~\cite{Banfi:2010xy}:
$(\frac{1}{2}\mu_r, \frac{1}{2}\mu_f)$, $(\frac{1}{2}\mu_r, \mu_f)$, $(\mu_r, \frac{1}{2}\mu_f)$, $(\mu_r, 2\mu_f)$, $(2\mu_r, \mu_f)$, and $(2\mu_r, 2\mu_f)$.
The default choice is $\mu_{\rm r} = \mu_{\rm f} = \pt$.
These scale variations modify the prediction of the inclusive jet cross section by about $5\%$ to $10\%$.
Following the PDF4LHC recommendation~\cite{Alekhin:2011sk}, PDF uncertainties are evaluated via a prescribed envelope, defined as the maximum variation between different PDF sets constructed from CT10~\cite{Lai:2010vv}, MSTW2008NLO~\cite{Martin:2009iq}, and NNPDF2.0~\cite{Ball:2010de}, including their respective uncertainties and using their respective default values of the strong coupling constant $\alpha_s(m_Z)$ = 0.1180, 0.1190, and 0.1202. The middle of the envelope is taken as the central prediction. The uncertainties are on the order of $10\%$ up to a $\pt$ of $800\GeV$, except when approaching the kinematic limit where they can be as large as 40\%.
Finally, an additional uncertainty from the current knowledge of the strong coupling constant is calculated from the CT10as PDF set~\cite{Lai:2010vv} with values of $\alpha_s(m_Z)$ varied conservatively by $\pm0.002$ and added in quadrature to the PDF uncertainty. The uncertainties due to these variations in $\alpha_s(m_Z)$ are between 2.5 to 5.0\%.
The PDF uncertainties are dominated by differences between PDF sets in the PDF4LHC recommendation for 50~$<\pt<$~500~GeV, and by uncertainties within a single PDF set for $\pt > 500$~GeV.

The fully-corrected inclusive jet cross section is shown in Figs.~\ref{fig:UnsmearedSpectra} and \ref{fig:UnsmearedRatio}.  Figure~\ref{fig:UnsmearedSpectra} shows the jet \pt spectra between 18 and 1100~GeV, falling over 10 orders of magnitude in rate, and for six different rapidity bins.
The comparison with the theoretical NLO prediction, corrected for NP effects, is more easily discerned in Fig.~\ref{fig:UnsmearedRatio}, which provides the ratio of the jet \pt spectra from data to the theoretical prediction for each of the six rapidity bins.  The total theoretical systematic uncertainty from the prediction is superimposed as solid lines above and below unity, and the total systematic uncertainty due to experimental effects is centered on the data points as a shaded band.
The central predictions for the CT10, MSTW2008NLO, and NNPDF2.0 PDF sets are also overlayed. The PDF uncertainties are large and asymmetric at high jet \pt, dominating the theoretical uncertainty band. Nevertheless, compared to the PDF4LHC recommendation, similar trends between data and the central prediction of each PDF set are observed.
Within the experimental and theoretical uncertainties, the predictions are seen to be consistent with the data across a wide range of jet \pt and rapidities, although the predictions are systematically above the data.

In conclusion, using a data sample corresponding to $34$~pb$^{-1}$ of integrated luminosity from \Pp\Pp\  collisions recorded by the CMS detector at the LHC with a center-of-mass energy of 7~TeV, the jet transverse momentum spectrum has been measured for $18 < \pt < 1100\GeV$ and for six rapidity bins up to $|y| = 3.0$. %The kinematic range of this study extends to the highest values of jet \pt ever observed.
The dominant systematic uncertainties arise from the absolute jet momentum scale and resolution, as well as the integrated luminosity measurement.  The NLO pQCD predictions for the inclusive jet cross section, corrected for non-perturbative effects and using the PDF4LHC recommendations, are generally in agreement with the data.  This measurement extends to the highest values of jet \pt ever observed.

We wish to congratulate our colleagues in the CERN accelerator departments for the excellent performance of the LHC machine. We thank the technical and administrative staff at CERN and other CMS institutes, and acknowledge support from: FMSR (Austria); FNRS and FWO (Belgium); CNPq, CAPES, FAPERJ, and FAPESP (Brazil); MES (Bulgaria); CERN; CAS, MoST, and NSFC (China); COLCIENCIAS (Colombia); MSES (Croatia); RPF (Cyprus); Academy of Sciences and NICPB (Estonia); Academy of Finland, ME, and HIP (Finland); CEA and CNRS/IN2P3 (France); BMBF, DFG, and HGF (Germany); GSRT (Greece); OTKA and NKTH (Hungary); DAE and DST (India); IPM (Iran); SFI (Ireland); INFN (Italy); NRF and WCU (Korea); LAS (Lithuania); CINVESTAV, CONACYT, SEP, and UASLP-FAI (Mexico); PAEC (Pakistan); SCSR (Poland); FCT (Portugal); JINR (Armenia, Belarus, Georgia, Ukraine, Uzbekistan); MST and MAE (Russia); MSTD (Serbia); MICINN and CPAN (Spain); Swiss Funding Agencies (Switzerland); NSC (Taipei); TUBITAK and TAEK (Turkey); STFC (United Kingdom); DOE and NSF (USA).

\pagebreak

\clearpage
\bibliography{auto_generated}

\clearpage

\cleardoublepage\appendix\section{The CMS Collaboration \label{app:collab}}\begin{sloppypar}\hyphenpenalty=5000\widowpenalty=500\clubpenalty=5000\textbf{Yerevan Physics Institute,  Yerevan,  Armenia}\\*[0pt]
S.~Chatrchyan, V.~Khachatryan, A.M.~Sirunyan, A.~Tumasyan
\vskip\cmsinstskip
\textbf{Institut f\"{u}r Hochenergiephysik der OeAW,  Wien,  Austria}\\*[0pt]
W.~Adam, T.~Bergauer, M.~Dragicevic, J.~Er\"{o}, C.~Fabjan, M.~Friedl, R.~Fr\"{u}hwirth, V.M.~Ghete, J.~Hammer\cmsAuthorMark{1}, S.~H\"{a}nsel, M.~Hoch, N.~H\"{o}rmann, J.~Hrubec, M.~Jeitler, W.~Kiesenhofer, M.~Krammer, D.~Liko, I.~Mikulec, M.~Pernicka, H.~Rohringer, R.~Sch\"{o}fbeck, J.~Strauss, A.~Taurok, F.~Teischinger, P.~Wagner, W.~Waltenberger, G.~Walzel, E.~Widl, C.-E.~Wulz
% ...
\vskip\cmsinstskip
% ...
1:~~Also at CERN, European Organization for Nuclear Research, Geneva, Switzerland\\
% ...

\end{sloppypar}
\end{document}